\documentclass[twocolumn,showpacs,preprintnumbers,amsmath,amssymb]{revtex4}
\usepackage{graphicx}% Include figure files
\usepackage{dcolumn}% Align table columns on decimal point
\usepackage{bm}% bold math

\begin{document}

\title{Fano resonance in electronic transport through a quantum wire with a side-coupled quantum dot:
X-boson treatment}

\author{R. Franco}
\email{rfranco@if.uff.br}
\author{M. S. Figueira}
\email{figueira@if.uff.br} \affiliation{ Instituto de F\'{\i}sica,
Universidade Federal Fluminense (UFF), Av. Litor\^{a}nea s/n,
24210-340 Niter\'oi, Rio de Janeiro, Brasil, Caixa Postal 100.093}

\author{E. V. Anda}
\email{anda@fis.puc-rio.br} \affiliation{ Departamento de
F\'{\i}sica, Pontificia Universidade Cat\'olica do Rio de Janeiro,
22452-970 Rio de Janeiro, Brasil, Caixa Postal 38071}

\date{\today}
\begin{abstract}
The transport through a quantum wire with a side coupled quantum
dot is studied. We use the X-boson treatment for the Anderson
single impurity model in the limit of $U=\infty $. The conductance
presents a minimum for values of $T=0$ in the crossover from
mixed-valence to Kondo regime due to a destructive interference
between the ballistic channel associated with the quantum wire and
the quantum dot channel. We obtain the experimentally studied Fano
behavior of the resonance. The conductance as a function of
temperature exhibits a logarithmic and universal behavior, that
agrees with recent experimental results.
\end{abstract}

\pacs{73.63.Kv, 73.63.-b, 85.35.Ds }
 \maketitle

\section{Introduction}

\label{Sec1}

Quantum dots (QD) are small droplets of electrons confined in the
three spatial directions. In these systems the charge and  energy
are quantized as it occurs in natural atoms. The electron
transport in this ``artificial atom'' nanodevice is a topic of
intense research. Experiments performed in a single electron
transistor (SET), \cite{Gores2000} and by scanning tunnelling
microscopy (STM) on a single magnetic impurity on a metallic
surface \cite{impu1}, showed that the Kondo resonance, predicted
theoretically in these systems \cite{Kondo1} and experimentally
measured \cite {Kondo2}, is present simultaneously with a Fano
resonance. The Fano resonance appears when there is a quantum
interference process in a system consisting of a continuous
degenerated spectrum with a discrete level, both non-interacting.
The interference is produced among the electrons that circulate
along the two channels of the system constituted by the discrete
level and the continuous band. In general the device is designed
such that the current goes through the dot itself. Recently,
another configuration in which the dot is laterally linked to the
quantum wire has been studied,\cite{Kang2001}. This situation
mimics to some extent a metallic compound doped by magnetic
impurities. As in the problem of the metal, theoretical studies of
a dot laterally attached to a wire have shown that the Kondo
effect interferes with the transport channel reducing and
eventually eliminating the transmission of charge along it. A
similar configuration has been proposed where the dot is laterally
attached to a ring threaded by a magnetic field  \cite{Kondo4}.

Theoretically, these systems can be described by the Anderson
single impurity model (AIM). Structures with dots embedded or side
coupled to leads or interacting coupled quantum dots have been
studied using a variety of numerical and diagrammatic Green's
function techniques \cite{Kondo1,Torio2001}. The slave boson mean
field theory (SBMFT) \cite{Coleman84,SBMFT}, in particular, has
been applied to study these systems in the limit of the Coulomb
repulsion $U\rightarrow \infty $ and at low temperatures. This
approximation is attractive  because with a small numerical
effort, it is capable of qualitatively describes the Kondo regime,
although in a restricted region of the system parameter space.
Unfortunately, the SBMFT presents unphysical second order phase
transitions outside this region. The impurity decouples from the
rest of the system when $T>T_{K}$, where $T_{K}$ is the Kondo
temperature, or when $\mu >> E_{f,\sigma }$, where $\mu $ is the
chemical potential  and $E_{f,\sigma }$ is the energy of the
localized level \cite {Coleman85,X-boson2}. To circumvent these
problems and maintaining the simplicity of the calculation and the
ideas involved, recently we introduced the X-boson method
\cite{X-boson2,X-boson1,FisicaA}, inspired in the slave boson
formalism. We solve the problem of non-conservation of probability
(completeness) using the chain cumulant Green's functions
\cite{FFM,Infinite,Ufinito,ChainPhi}. These ideas were used to
solve the AIM and the Periodic Anderson Model. We remove the
spurious phase transitions of the SBMFT, which permits to study
the system for a range of temperatures that includes the $T>T_{K}$
region.

\begin{figure}[ht]
\includegraphics[clip,width=0.45\textwidth, height=0.25\textheight
,angle=0.0]{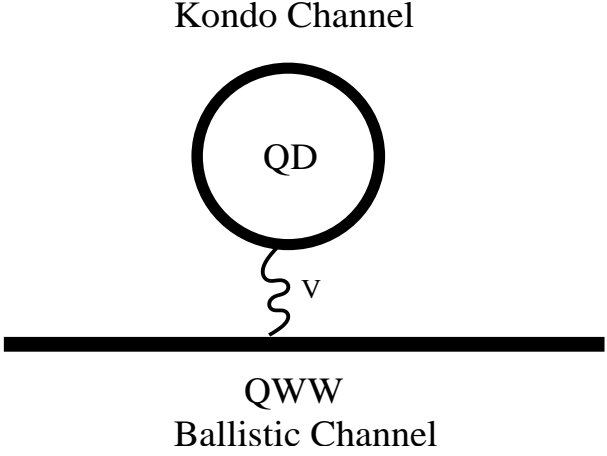} \caption{Pictorial representation of the
quantum well wire (QWW) coupled via hybridization (V) with  a side
quantum dot (QD).} \label{QDfig}
\end{figure}

In this work we apply the X-boson method for the single impurity
case to describe the transport problem through a quantum wire with
a side coupled QD as represented in Fig. \ref{QDfig}, in the limit
of $U\rightarrow \infty$, without restriction in the temperature.

Our results are in good agreement with experimental measurements,
in particular we reproduce the Fano behavior of the resonance and
the logarithmic dependence of the conductance dip amplitude as a
function of temperature \cite{Gores2000} in a SET and we obtain
for the conductance as function of temperature a logarithmic and
universal behavior \cite{Kondo2,Italiano2002}. We restrict our
analysis to systems that are not in the extreme Kondo regime
because the X-boson approach, in its present form, follows closely
the Friedel sum rule in the mixed-valence and in the moderate
Kondo regime, but fails to fulfil it in the extreme Kondo regime,
when $E_{f,\sigma}$ is far below the chemical potential ($\mu=0$
in our case).

\bigskip

\section{ Model, Method and Conductance}

\label{Sec2}

The model we use to describe the system is the Anderson impurity Hamiltonian
in the $U=\infty $ limit, employing the Hubbard $X$ operators to project out
the states with double occupation on the QD. We obtain

\begin{eqnarray}
H &=&\sum_{\mathbf{k},\sigma }E_{\mathbf{k},\sigma
}c_{\mathbf{k},\sigma }^{\dagger }c_{\mathbf{k},\sigma
}+\sum_{\sigma }\ E_{f,\sigma }X_{f,\sigma
\sigma }  \nonumber \\
&&+\sum_{\mathbf{k},\sigma}\left( V_{f,\mathbf{k},\sigma }X_{f,0\sigma
}^{\dagger }c_{\mathbf{k},\sigma }+V_{f,\mathbf{k},\sigma }^{\ast }c_{%
\mathbf{k},\sigma }^{\dagger }X_{f,0\sigma }\right) .  \label{Eq.3}
\end{eqnarray}

\noindent The first term of the equation represents the
Hamiltonian of the conduction electrons ($c$-electrons),
associated with the wire. The second term describes the QD and the
last one corresponds to the interaction between the $c$-electrons
and the QD. This Hamiltonian can be treated by the X-boson
cumulant approach \cite{X-boson2,X-boson1,FisicaA} for the
impurity case.

At low temperature and bias voltage, electron transport is
coherent and a linear-response conductance is given by the
Landauer-type formula \cite{Kang2001}

\begin{equation}
G=\frac{2e^2}{h}\int{\left(-\frac{\partial n_{F} }{\partial \omega}%
\right)S(\omega) d\omega} ,  \label{Landauer}
\end{equation}

\noindent where $n_{F}$ is the Fermi distribution function and
$S(\omega )$ is the transmission probability of an electron with
energy $\hbar \omega $, given by,

\begin{equation}
S(\omega)=\Gamma^{2}\vert{G_{QD,\sigma}}\vert^{2} ,  \label{Transmission}
\end{equation}

\noindent where $\Gamma=V^{2}/\Delta$ is the coupling between the
QD and the wire, with $\Delta =\frac{\pi V^{2}}{2D}$ and
$G_{QD,\sigma }(\omega )$ is the Green function at the site of the
QD with spin $\sigma .$ It can be written in terms of the Green's
functions of the localized level $G_{f}^{\sigma }(\omega )$ and
the conduction electrons, $G_{c}^{\sigma }(\omega ).$ Supposing
that the interaction between the dot and the rest of the system is
independent of $\mathbf k$ and $\sigma ,$ $V=V_{f,\mathbf
{k},\sigma }$, the Green function at the QD can be written as,

\begin{equation}\label{Gqd}
G_{QD,\sigma }=G_{c}^{\sigma} V G_{f}^{\sigma } V G_{c}^{\sigma}+
G_{c}^{\sigma} .
\end{equation}

In the chain X-boson method, the cumulant Green's function,
considering a constant density of states for the wire, $-D\leq
\varepsilon_{\bf{k}} \leq D $ are given by
\cite{X-boson2,X-boson1,FisicaA}

\begin{equation}\label{Gff}
G_{f}^{\sigma }(z)=\frac{-D_{\sigma}}{z-\tilde{E_{f}}-\frac{V^{2}
D_\sigma}{2D}ln\left| \frac{z+D}{z-D}\right|} ;
\end{equation}

\begin{equation}\label{Gcc}
G_{c}^{\sigma } (z)=-\frac{1}{2D}ln\left|\frac{z+D}{z-D}\right| ;
\end{equation}

\begin{equation}\label{Gfc}
G_{fc}^{\sigma }(z)=-\frac{\frac{-V D_\sigma}{2D}ln\left|
\frac{z+D}{z-D}\right|} {z-\tilde{E_{f}}-\frac{V^{2}
D_\sigma}{2D}ln\left| \frac{z+D}{z-D}\right|} ,
\end{equation}

\noindent where $z=\omega +i\eta $, the quantity $D_{\sigma
}=\left\langle X_{0,0}\right\rangle +n_{f,\sigma }$ is responsible
for the correlation in the chain X-boson approach and lead to
essential differences with the uncorrelated case obtained using
the slave-boson method \cite {X-boson2} and the  conduction Green
function $G_{c}^{\sigma } (z)$ represents a ballistic channel. The
occupation number of the quantum dot is given by
$n_{QD}=2n_{f,\sigma }$, where $n_{f,\sigma }$ is the occupation
number of the electrons with spin $\sigma $ in the QD and
$\tilde{E_{f}} =E_{f}+\Lambda $, where $\Lambda $ is a parameter
of the X-boson method given by \cite{X-boson2}

\begin{eqnarray}
\Lambda &=&\frac{-V^{2}}{D}\int_{-\infty }^{\infty }d\omega \hspace{0.1cm}%
\rho _{f}(\omega )\ln \left| \frac{\omega +D}{\omega -D}\right|  \nonumber \\
&&\times \frac{(\omega ^{2}-D^{2})\hspace{0.1cm}n_{F}(\omega )}{(\omega
^{2}-D^{2}+V^{2}D_{\sigma })},  \label{Lambda1}
\end{eqnarray}

\noindent where $\rho _{f}(\omega )=\sum_{i}\delta (\omega -\omega _{i})$ is
the density of states at the $QD$ and $\omega _{i}$ are the poles of the $%
G_{f}^{\sigma }(\omega )$.

\section{Results and Discussion}

\label{Sec4}

\begin{figure}[ht]
\includegraphics[clip,width=0.40\textwidth, height=0.35\textheight%
,angle=-90.]{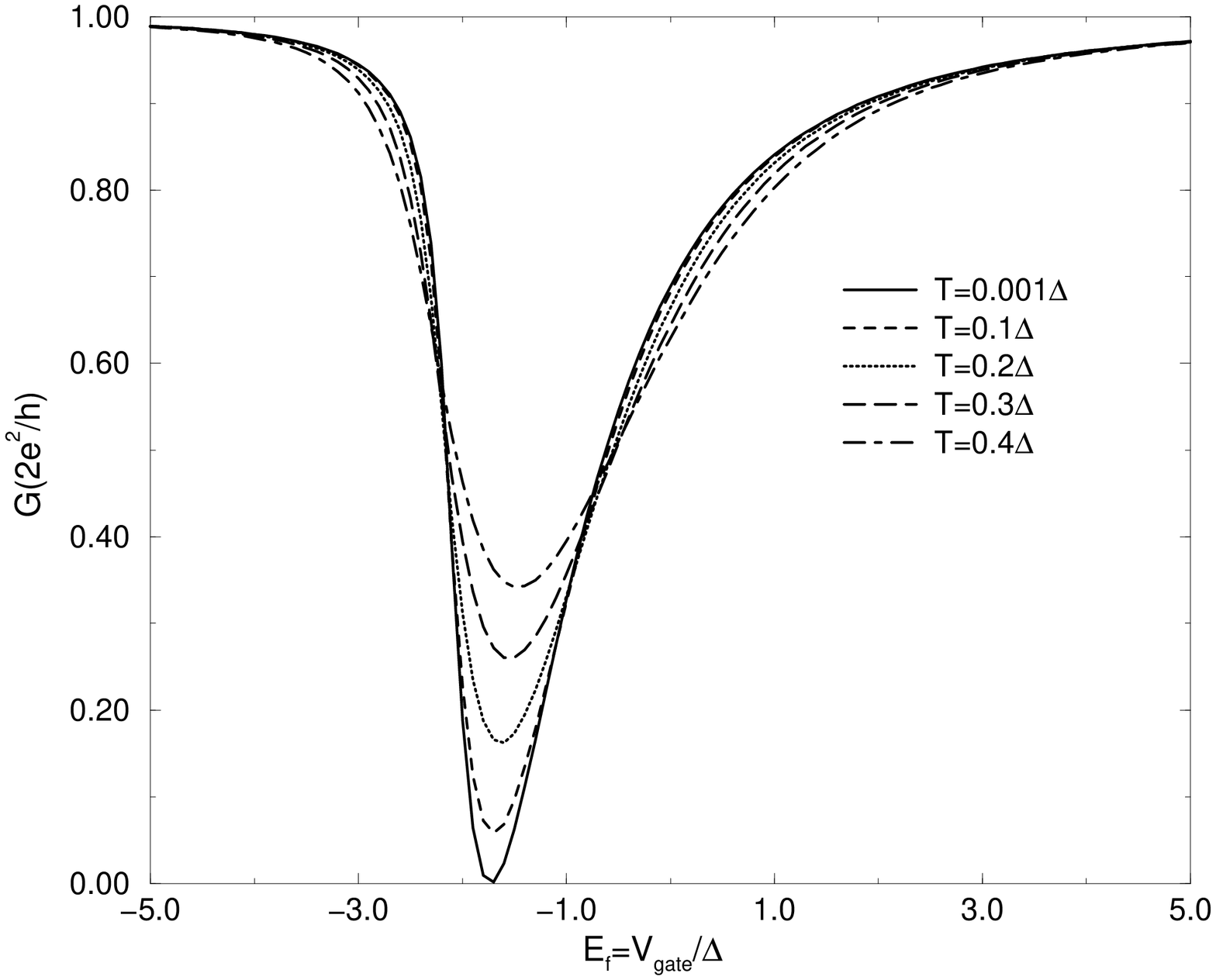} \caption{Conductance $G$ vs
$E_{f}=V_{gate}/\Delta$ for different temperatures (low
temperatures).} \label{fig1_QD}
\end{figure}

\begin{figure}[ht]
\includegraphics[clip,width=0.40\textwidth, height=0.35\textheight%
,angle=-90.]{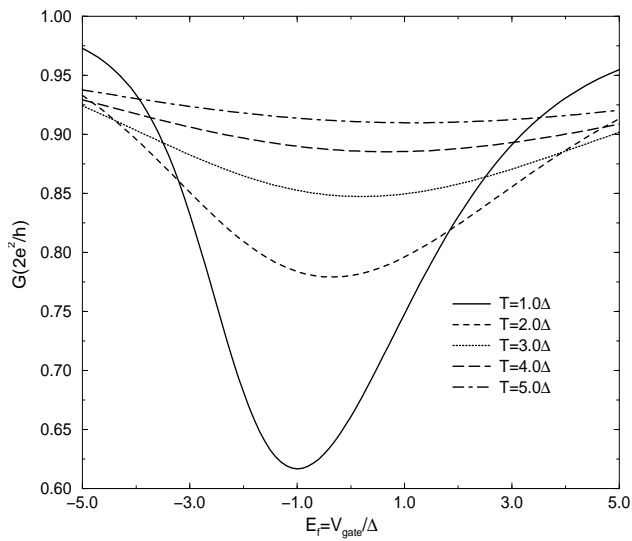} \caption{Conductance $G$ vs
$E_{f}=V_{gate}/\Delta$ for different temperatures (high
temperatures).} \label{fig2_QD}
\end{figure}

The Figs. \ref{fig1_QD},\ref{fig2_QD} show the conductance in
units of $2e^{2}/h$, as function of the gate voltage $V_{gate}$,
given in units of $\Delta$, for different temperatures. We
represent the low and high temperature regions in Fig.
\ref{fig1_QD} and \ref{fig2_QD} respectively. In all the cases we
consider $D=100\Delta $ and $\mu=0.0$. The gate potential,
$V_{gate}$ is controlled experimentally \cite{Gores2000,Kondo2}
and allow us to modify the energy position level $E_{f}$ of the
quantum dot ($E_{f}=V_{gate}$), which is renormalized by the
$\Lambda $ parameter ($\tilde{E_{f}}=V_{gate}+\Lambda $) according
to the X-boson method. The conductance possesses an asymmetric
Fano resonance shape, which agrees well with theoretical and
experimental results \cite{Gores2000,Torio2001,Bulka2001}. We
obtained similar results to the one obtained by the SBMFT,
\cite{Kang2001}, at intermediate to low temperatures region
$T<T_{K}$ as can be seen from fig. \ref{fig1_QD}. In Fig.
\ref{fig2_QD}, we present the conductance for $T>T_{K}$ as a
function of the gate voltage for different temperatures. In this
region the SBMFT conductance is incorrect due to the spurious
second order phase transition associated with this method which
decouples the $QD$ from the rest of the system.

\begin{figure}[ht]
\includegraphics[clip,width=0.40\textwidth, height=0.35\textheight
,angle=-90.]{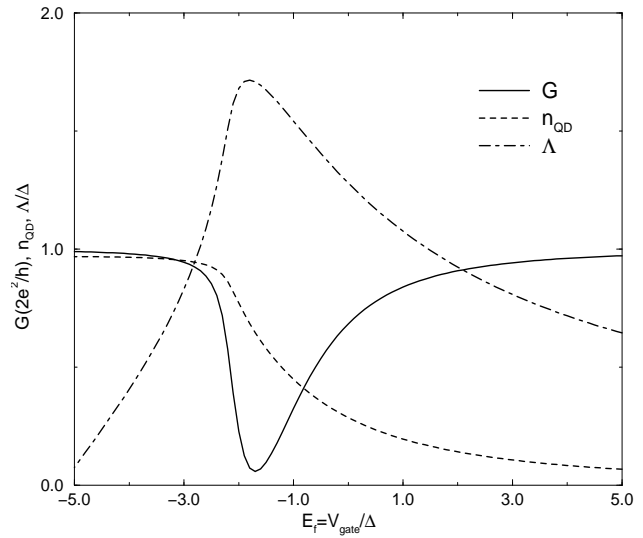} \caption{Conductance $G$, charge in the
QD, $n_{QD}$ and $\Lambda/\Delta$ parameter vs
$E_{f}=V_{gate}/\Delta$ at $T=0.1\Delta$.} \label{fig4_QD}
\end{figure}

\begin{figure}[ht]
\includegraphics[clip,width=0.40\textwidth, height=0.35\textheight%
,angle=-90.]{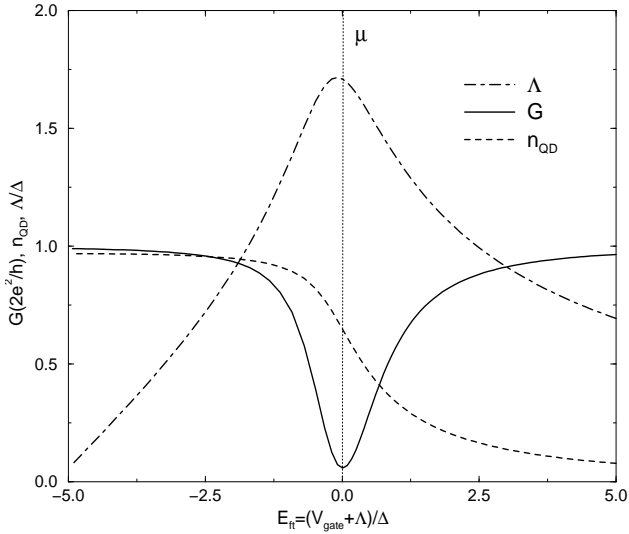} \caption{X-boson parameter $\Lambda$,
Conductance $G$ and charge in the QD, $n_{QD}$ vs
$\tilde{E_{f}}=(V_{gate}+\Lambda)/\Delta$ at $T=0.1\Delta$.}
\label{fig5_QD}
\end{figure}

In Fig. \ref{fig4_QD} we present at $T=0.1\Delta $, the
$\Lambda/\Delta$ parameter, the occupation number $n_{QD}$ (charge
in the QD) and the conductance vs $V_{gate}/\Delta$. The minimum
of the conductance and the maximum of the $\Lambda $ parameter
correspond to the same value of $V_{gate}$. The asymmetric shape
of the curves in Fig. \ref{fig4_QD}, as mentioned above, is a
result of the Fano behavior of the resonance. When the dot is in
the Kondo regime where $\tilde{E_{f}}\approx \mu ,$ the electron
has a channel to get through the system, going up and down to the
dot. This trajectory interferes with the path the electrons take
when they go straight along the leads, without visiting the dot,
giving rise to a Fano shape. In Fig. \ref {fig5_QD} we display the
same results shown in \ref{fig4_QD} as a function of
$\tilde{E_{f}}=(V_{gate}+\Lambda)/\Delta $. In contrast with the
Fig. \ref{fig1_QD}, we obtain a symmetric curve for the
conductance, which shows that the Fano behavior of the resonance
is associated with the renormalization of the parameters of the
system due to the Kondo effect.

\begin{figure}[ht]
\includegraphics[clip,width=0.40\textwidth, height=0.35\textheight%
,angle=-90.]{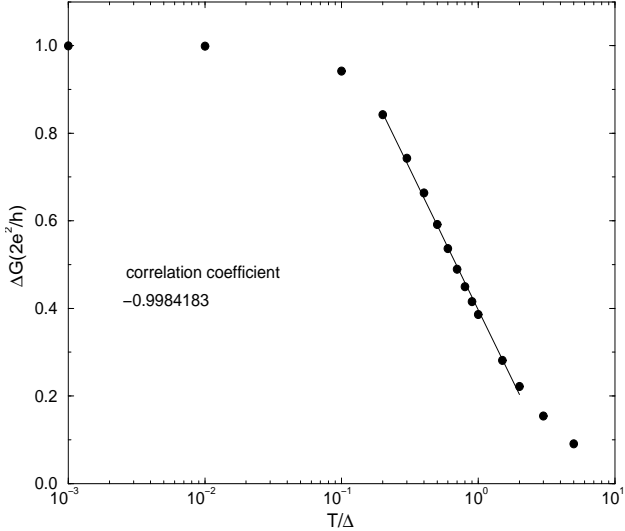} \caption{Dip amplitude conductance
$\Delta G$ measured from the background conductance (the
``distance'' between the minimum of the curve and the conductance
value $G=1.0$), as function of $T/\Delta$ (see Figs.
\ref{fig1_QD},\ref{fig2_QD}).} \label{fig6_QD}
\end{figure}

In Fig. \ref{fig6_QD} we show the minimum conductance amplitude as
a function of temperature, in units of $\Delta$. The logarithmic
behavior  presented in the interval between $0.2\Delta < T <
2.0\Delta $, agrees well with the experimental results obtained by
Gores et al in a SET \cite {Gores2000} (with $\Delta \approx
1.0K$) and reflects the crossover from the mixed-valence to the
Kondo regime, where the X-boson is a reliable approximation.

\begin{figure}[ht]
\includegraphics[clip,width=0.40\textwidth, height=0.35\textheight%
,angle=-90.]{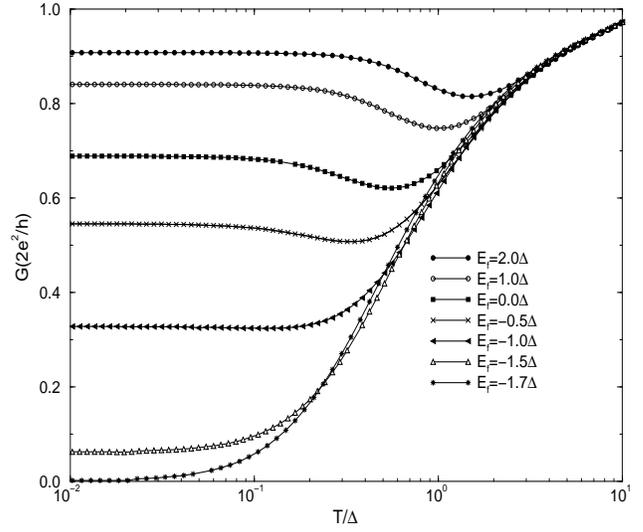}
\caption{Conductance $G$ as a function of $T/\Delta$, at different values of $%
E_{f}=V_{gate}/\Delta$, (crossover from the quasi-empty QD to the
Kondo regime).} \label{fig16_QD}
\end{figure}

In Fig. \ref{fig16_QD} we present the conductance $G$ vs
temperature $T/\Delta$, for different values of
$E_{f}=V_{gate}/\Delta.$ We describe the gradual crossover from
the quasi-empty quantum dot to the Kondo regime. The conductance
exhibits a minimum at high temperature that moves to the low
temperature region as $\tilde{E_{f}}$ approaches the chemical
potential level $\mu =0$. This minimum is associated with the
energy required to excite an electron from the chemical potential
energy $\mu $ up to the level $\tilde{E_{f}}$. These results are
in qualitative agreement with a recent theoretical calculation
\cite{Costi2001}, applying Wilson numerical renormalization group
approach (NRG) for a embedded QD in a quantum wire, taking into
account that the minimum in our case corresponds to the maximum in
this work.

\begin{figure}[ht]
\includegraphics[clip,width=0.40\textwidth, height=0.35\textheight%
,angle=-90.]{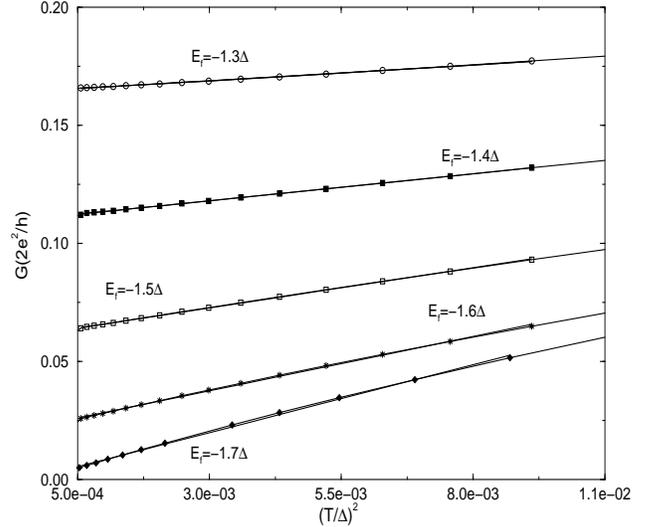} \caption{Conductance $G$ as a function of
$(T/\Delta)^{2}$, at different values of $E_{f}=V_{gate}$. The
linear behavior of the curves agrees well with the expected Fermi
liquid behavior from the mixed valence to the Kondo regime.}
\label{fig18_QD}
\end{figure}

The Fig. \ref{fig18_QD} shows, at low temperatures ($T \ll
T_{K}$), the conductance G vs $(T/\Delta)^2$, for values of
$E_{f}$ that corresponds to the crossover from the mixed valence
to the Kondo regime. The linear behavior of the curves agree well
with the expected Fermi liquid behavior, that can be represented
by the equation \cite{Italiano2002}

\begin{equation}
G\simeq G_{min}\left(1+\alpha \left(\frac{T}{T_{K}}\right)^{2}\right) ,
\label{Landau}
\end{equation}

\noindent where $\alpha $ is a parameter and $G_{min}$ is the
conductance at $T=0$. The different slopes of the straight lines
obtained reflect the different values of $T_{K}$ for each $E_{f}$.

\begin{figure}[ht]
\includegraphics[clip,width=0.37\textwidth, height=0.37\textheight%
,angle=-90.]{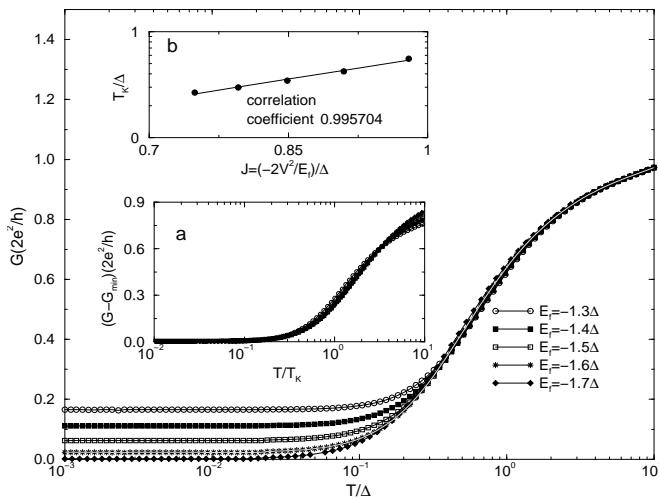} \caption{Conductance $G$ as a function of
$T/\Delta$, at different values of $E_{f}=V_{gate}/\Delta$. In the
inset $a$ we show the universal behavior of the crossover region
from the mixed valence to the Kondo regime; in the inset $b$ we
show the exponential dependence of $T_{K}$ as a function of
$J=-2V^2/E_{f}$.} \label{fig7b_QD}
\end{figure}

The Fig. \ref{fig7b_QD} presents the Conductance $G$ vs
$T/\Delta.$ It shows a minimum at low temperatures and a
logarithmic and universal behavior for intermediate temperatures.
This logarithmic evolution is a manifestation of the crossover
from the mixed-valence to the Kondo regime, as a consequence of
the renormalization of the localized level as the temperature is
varied. The deviation from the logarithmic behavior is associated
with the beginning of the high temperatures limit. The inset $a$
shows $G-G_{min}$ vs $\frac{T}{T_{K}}$ revealing the universal
behavior of the conductance at the Kondo regime while the inset
$b$ present the expected exponential behavior for the Kondo
temperature $T_{K}$ as function of $J=\frac{-2V^2}{E_{f}}$. The
values of $T_{K}$  are obtained from the straight lines slopes of
the Fig. \ref{fig18_QD} taking for the parameter $\alpha$ of Eq.
(\ref{Landau}), the value that results from adopting the Lacroix's
definition of the Kondo temperature \cite{Lacroix} for the case
$E_{f}=-1.7\Delta $ ($T_{K}$ is the temperature that corresponds
to the minimum of $d<c_{i\sigma}^{\dagger}f_{i\sigma}>/dT$). The
universal behavior agrees with experimental and theoretical
results for an embedded QD \cite {Kondo2,Italiano2002}, taking
into account that for this configuration, the maximum of $G$
corresponds in our case to a minimum.

\begin{figure}[ht]
\includegraphics[clip,width=0.40\textwidth, height=0.35\textheight,angle=-90.]
{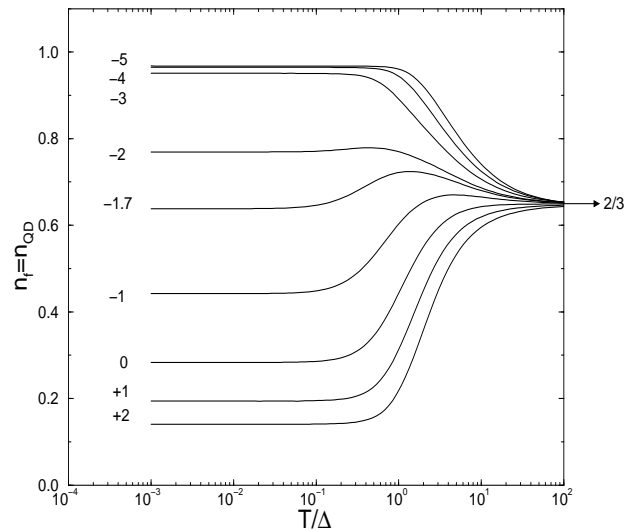} \caption{Charge in the QD, $n_{QD}$ as a function of
$T/{\Delta}$, at different values of $E_{f}=V_{gate}/\Delta$, for
all regimes of the model.} \label{fig15_QD}
\end{figure}

In Fig. \ref{fig15_QD} we present the charge in the QD as a
function of $T/\Delta$ for all regimes of the model, at different
values of $E_{f}=V_{gate}/\Delta$. We can compare the X-boson
results with the very accurate results obtained
\cite{Costi94,Costi96} using a numerical renormalization group
(NRG) calculation. The X-boson charge in the quantum dot exhibits
the same shape as obtained by this method (see their Fig. 6). At
high temperatures, in all cases, the charge goes to the correct
high temperature limit of $2/3$.

These results permit to consider the X-boson approximation as a
quantitative correct method in the empty dot regime and in the
crossover from the mixed valence to the Kondo regime, for all
temperatures and for all values of the parameters of the model,
although it is not reliable in the extreme Kondo limit (when
$E_{f,\sigma}$ is far below the chemical potential $\mu=0$), where
the Friedel sum rule \cite{Fridel} is not fulfilled. This behavior
makes this method complementary to the Non Crossing Approximation
(NCA) that is satisfactory in the extreme Kondo limit but does not
behaves adequately along the crossover from the mixed valence to
the empty dot regime at low temperatures. This is a consequence of
the fact that in the NCA, the Kondo resonance survive even in the
empty orbital regime producing a spurious high density of states
at the Fermi level \cite{Costi96}.

\section{Summary and Conclusions}

\label{Sec5}

We have calculated the conductance for a quantum wire with a side
coupled QD as a function of gate voltage and temperature, for
$U=\infty $. Our results agree well with recent experimental
studies \cite {Gores2000,Kondo2}. We obtain the Fano shape for the
conductance, its suppression at the Kondo regime at $T=0$, due to
destructive interferences between the Kondo channel (QD) and the
conduction channel (quantum wire) and the logarithmic behavior of
the conductance dip amplitude as function of temperature as
reported in a recent experiment \cite {Gores2000}. This behavior
is a manifestation of the crossover from the mixed-valence to the
Kondo regime and is valid for the conductance within this region.
These results agree with experimental works for a QD embedded in a
quantum wire \cite{Kondo2}. We obtain  a universal behavior of the
conductance as a function of temperature for different values of
$E_{f}$ that as well agrees with recent experimental results
\cite{Kondo2} for a embedded QD configuration. We report novel
results for the conductance as a function of temperature, below
and above $T_{K}$ for a lead with a side coupled QD
configuration.We were able to show that the X-boson approximation
is a simple and appropriate tool to study mesoscopic transport
including quantum dots in the Kondo regime.
\begin{acknowledgments}
We acknowledge Profs. M. E. Foglio (Unicamp) and A. A. Aligia
(CAB-Bariloche) for helpful discussions and the financial support from
the Rio de Janeiro State Research Foundation (FAPERJ) and National
Research Council (CNPq).
\end{acknowledgments}

\end{document}